\documentclass[12pt]{article}
\newlength{\mytopmargin}
\newlength{\myleftmargin}
\def\zz{\rlx\hbox{\small \sf Z\kern-.4em Z}}

\begin{document}
\vspace{1cm}
\noindent
\begin{center}{\large \bf  Exact Wigner surmise type
evaluation of the spacing distribution
in the bulk of the scaled random matrix ensembles}
\end{center}
\vspace{5mm}

\begin{center}
 P.J.~Forrester and N.S.~Witte$^*$
\\
\it Department of Mathematics and Statistics, $^*$(\& School of Physics),
\\
University of Melbourne, Parkville, Victoria
3052, Australia
\end{center}
\vspace{.5cm}

\small
\begin{quote}
Random matrix ensembles with orthogonal and unitary symmetry correspond to
the cases of real symmetric and Hermitian random matrices respectively.
We show that
the probability density function for the corresponding spacings between
consecutive eigenvalues 
can be written exactly in the Wigner surmise type form $a(s) e^{-b(s)}$
for $a$ simply related to a Painlev\'e transcendent and $b$ its 
anti-derivative. A formula consisting of the sum of two such terms is
given for the symplectic case (Hermitian matrices with real quaternion
elements).
\end{quote}


\vspace{.5cm}
It is  well established that universal features of the spectrum of classically
chaotic quantum systems are correctly described by random matrix ensembles of 
an appropriate symmetry \cite{Po65,BFPW81,Me91}. Generically there are three
symmetry classes corresponding to two distinct time reversal symmetries
plus the situation in which time reversal symmetry is absent. The level
repulsion --- which is a characteristic of the spectra of chaotic systems
and is not present in the spectra of integrable systems --- differs for
each of the symmetry classes. Thus let $p(s)$ denote the probability density
function for the spacing between consecutive levels, calculated after
the energy levels are first rescaled so that the mean spacing is unity. 
Then $p(s) \propto s$ for a time reversal symmetry $T$ such that $T^2=1$,
$p(s) \propto s^2$ in the absence of time reversal symmetry while
$p(s) \propto s^4$ for a time reversal symmetry such that $T^2 = -1$. It
is therefore convenient to distinguish the three cases by the label
$\beta$ and so write $p_\beta(s)$, where $\beta = 1,2$ or 4 depending on the
small $s$ behaviour of $p_\beta(s)$. In this work succinct expressions
for $p_\beta(s)$ will be given in terms of Painlev\'e transcendents.

The full probability density function  $p_\beta(s)$ is by far the most studied
statistic in relation to empirical data. For example
in the early work \cite{RP60}
(this article is reprinted in \cite{Po65})
on the energy levels of complex nuclei one sees in Figure 3 an empirical
bar graph of $p_1(s)$ obtained from experimental data plotted on
the same graph as a theoretical approximation to $p_1(s)$
known as the Wigner surmise. This approximation is given by the functional
form $p_1^{W}(s) = (\pi s / 2) e^{-(\pi s /2)^2}$. The fact that the Wigner
surmise is an approximation rather than exact was soon realized
\cite{Me60}, and the task of calculating the exact form of $p_1(s)$
was undertaken \cite{Ga61}. Actually the quantity calculated was
$E_\beta(0;s)$, the probability there are no eigenvalues in an interval
of length $s$. One then computes $p_\beta(s)$ using numerical differentiation
via the formula
\begin{equation}\label{ws2}
p_\beta(s) = {d^2 \over ds^2} E_\beta(0;s).
\end{equation}
The quantity $E_1(0;s)$ was tabulated by first obtaining a formula involving the
eigenfunctions and eigenvalues of a certain integral operator. The same
general approach can be taken to compute $p_\beta(s)$ for $\beta = 2$
and $\beta = 4$ \cite{Me91}. The work \cite{Ga61} revealed that the
Wigner surmise, although an excellent approximation, can be in error by
up to 2\% from the exact value.

There is a celebrated example of the empirical determination of $p_2(s)$
which is so accurate that the exact value to an accuracy of three to four
decimal places is essential. This occurs in Odlyzko's numerical computation
\cite{Od87,Od89}
of the large zeros of the Riemann zeta function on the critical line, in
particular zero number $10^{20}$ and $10^7$ of its neighbours, which
according to the Montgomery-Odlyzko law \cite{KS99} are conjectured to
have the same statistical properties as the eigenvalues of large
dimensional random matrices.

Nearly 20 years after the numerical computation of $E_1(0;s)$, Jimbo et
al.~\cite{JMMS80} computed the exact functional form of $E_2(0;s)$ in
terms of a Painlev\'e V transcendent. Thus with $\sigma(s)$ defined as the
solution of the nonlinear equation
$$
(s\sigma'')^2 + 4(s\sigma ' - \sigma)(s\sigma ' - \sigma + (\sigma ')^2)=0
$$
subject to the boundary condition $\sigma(s) \sim - s/\pi  - (s/\pi)^2$
as $s \to 0$, it was shown 
\begin{equation}\label{ws3}
E_2(0;s) = \exp \Big ( \int_0^{\pi s} {\sigma(t) \over t} \, dt \Big ).
\end{equation}
Furthermore, using a known inter-relationship between $E_2$ and $E_1$
\cite{Dy62c} a formula equivalent to
\begin{equation}\label{ws4}
E_1(0;s) = \Big ( E_2(0;s) \Big )^{1/2}
\exp \bigg ( {1 \over 2} \int_0^{\pi s}
\Big ( - {d \over dx} {\sigma(x) \over x} \Big )^{1/2} \, dx \bigg )
\end{equation}
was presented. With $E_2$ and $E_1$ determined, $E_4$ can be computed from
the formula \cite{MD63}
\begin{equation}\label{ws5}
E_4(0;s/2) = {1 \over 2} \bigg ( E_1(0;s) + {E_2(0;s) \over E_1(0;s)} \bigg ).
\end{equation}

Thus the exact functional form of $p_\beta(s)$ can be obtained by substituting
(\ref{ws3}), (\ref{ws4}) and (\ref{ws5}) as appropriate in (\ref{ws2}) and
computing the second derivative. However the resulting formulas lack the 
aesthetic appeal of (\ref{ws3})--(\ref{ws5}), and from a practical
viewpoint have the drawback of requiring not only the computation of
$\sigma(x)$ but also its first and second derivative.

In this work we will show that for $\beta = 1$ and $\beta = 2$
the derivative
$$
{d \over ds} E_\beta(0;s)
$$
can be written in a form analogous to (\ref{ws3}), thus allowing 
expressions for $p_\beta(s)$ to be obtained which have the Wigner
surmise type structure $p_\beta(s) = a(s) e^{-b(s)}$.
A similar result will be obtained in the case $\beta = 4$.
The starting point for our
calculation in the case $\beta = 2$ is the formula (\ref{ws3}), but in the
case $\beta = 1$ we use a formula distinct from (\ref{ws4}). This latter
formula requires introducing a Painlev\'e transcendent $\sigma_{B}$
(the use of the subscript $B$ is motivated by the relation of this 
function to the Bessel kernel \cite{TW94b})
which satisfies the nonlinear equation
\begin{equation}\label{nw0}
(s \sigma_{B}'')^2 +
\sigma_{B}'(\sigma_{B} - s \sigma_{B}')(4 \sigma_{B}' - 1) -
{1 \over 4} (\sigma_{B}')^2 = 0
\end{equation}
subject to the boundary condition
\begin{equation}\label{nw0'}
 \sigma_{B}(s) \sim {s^{1/2} \over \pi} + {2 s\over \pi^2}.
\end{equation}
In terms of this function we have \cite{Fo99a}
\begin{equation}\label{nw1}
E_1(0;s) = \exp \bigg ( - \int_0^{(\pi s /2)^2}
{\sigma_{B}(x) \over x} \, dx \bigg ).
\end{equation}

Our objective is to express 
the derivatives of (\ref{ws3}) and (\ref{nw1}) in terms of 
functional forms analogous to the original expressions.
This is achieved by using
some mathematical results \cite{CS93} relating to certain second
order differential equations, of the second degree in $y''$, which possess
the Painlev\'e property. We find
the functional forms involve different Painlev\'e
transcendents to those occuring in (\ref{ws3}) and (\ref{nw1}).
In the case $\beta = 2$, the required Painlev\'e transcendent is specified
by the solution of the nonlinear equation
\begin{equation}\label{nw2}
s^2 (\tilde{\sigma}'')^2 + 4 (s \tilde{\sigma}' - \tilde{\sigma})
(s \tilde{\sigma}' - \tilde{\sigma} + (\tilde{\sigma}')^2) -
4(\tilde{\sigma}')^2 = 0
\end{equation}
subject to the boundary condition
$$
 \tilde{\sigma}(s) \sim - {s^3 \over 3 \pi}.
$$
For $\beta = 1$ the corresponding Painlev\'e transcendent is specified
by the solution of the nonlinear equation
\begin{equation}\label{nw3}
s^2(\tilde{\sigma}_{B}'')^2 =
( 4(\tilde{\sigma}_{B}')^2 - \tilde{\sigma}_{B}')
(s \tilde{\sigma}_{B}' - \tilde{\sigma}_{B}) +
{9 \over 4} (\tilde{\sigma}_{B}')^2 -
{3 \over 2}  \tilde{\sigma}_{B}' + {1 \over 4}
\end{equation}
subject to the boundary condition
\begin{equation}\label{bca}
\tilde{\sigma}_{B}(s) \sim {s \over 3} - {s^2 \over 45} +
{8 s^{5/2} \over 135 \pi}.
\end{equation}

In terms of the transcendents specified by (\ref{nw2}) and (\ref{nw3}) our
results are
\begin{equation}\label{a1}
{d \over ds} \exp \int_0^{\pi s} {\sigma(t) \over t} \, dt
= - \exp \int_0^{\pi s} {\tilde{\sigma}(t) \over t} \, dt
\end{equation}
\begin{equation}\label{a2}
{d \over ds} \exp \bigg ( - \int_0^{(\pi s/2)^2}
{\sigma_{B}(t) \over t} \, dt \bigg )
= - \exp \bigg ( - 
\int_0^{(\pi s/2)^2} {\tilde{\sigma}_{B}(t) \over t} \, dt \bigg ).
\end{equation}
Recalling (\ref{ws3}), (\ref{nw1}) and (\ref{ws2}) we therefore have
\begin{eqnarray}
p_2(s) & = & - {\tilde{\sigma}(\pi s) \over s}
\exp \int_0^{\pi s} {\tilde{\sigma}(t) \over t} \, dt \label{bw1} \\
p_1(s)  & = & {2 \tilde{\sigma}_{B}((\pi s /2)^2) \over s}
 \exp \bigg ( - 
\int_0^{(\pi s/2)^2} {\tilde{\sigma}_{B}(t) \over t} \, dt \bigg ).
\label{bw2}
\end{eqnarray}

In the case $\beta = 4$ we start with the formula \cite{Fo99a}
\begin{equation}
E_4(0;s) = {1 \over 2} \exp \Big ( - \int_0^{(\pi s)^2} {\sigma_B(x) \over
x} \, dx \Big ) +
{1 \over 2} \exp \Big ( - \int_0^{(\pi s)^2} {\sigma_{B+}(x) \over
x} \, dx \Big )
\end{equation}
where $\sigma_{B+}$ satisfies the same d.e.~(\ref{nw0}) as $\sigma_B$, but is
subject to the boundary condition
$$
\sigma_{B+}(s) \sim
{s^{3/2} \over 3 \pi} \Big ( 1 + {\rm O}(s) \Big ) +
{2 \over 3} \Big ( {1 \over 3 \pi} \Big )^2 s^3  \Big ( 1 + {\rm O}(s) \Big ).
$$
Proceeding as in the derivation of (\ref{a2}) we can derive the result
\begin{equation}
{d \over dx} \exp \bigg ( - \int_0^x {{\sigma}_{B+}(t) \over t} \,
dt \bigg ) = - {x^{1/2} \over 3 \pi}
 \exp \bigg ( - \int_0^x {\tilde{\sigma}_{B+}(t) \over t} \,
dt \bigg )
\end{equation}
where $\tilde{\sigma}_{B+}$ satisfies the differential equation
\begin{equation}
s^2( \tilde{\sigma}_{B+}'')^2 =
(4(\tilde{\sigma}_{B+}')^2 -  \tilde{\sigma}_{B+})(
s \tilde{\sigma}_{B+}' -  \tilde{\sigma}_{B+})
+ {25 \over 4} ( \tilde{\sigma}_{B+})^2 - {5 \over 2}
 \tilde{\sigma}_{B+} + {1 \over 4}
\end{equation}
subject to the boundary condition
$$
 \tilde{\sigma}_{B+}(s) \sim
{s \over 5} \Big ( 1 + {\rm O}(s) \Big ) + 
{8s^{7/2} \over 3^3 \cdot 5^3 \cdot 7\pi }
\Big ( 1 + {\rm O}(s) \Big ).
$$
We then have
\begin{equation}\label{15}
p_4(s) = 2 p_1(2s) + {2 \pi^2 s \over 3}
\Big ( \tilde{\sigma}_{B+}((\pi s)^2) - 1 \Big )
\exp \bigg ( - \int_0^{(\pi s)^2}
{ \tilde{\sigma}_{B+}(t) \over t} \, dt \bigg ).
\end{equation}

The formulas (\ref{bw1}), (\ref{bw2}) and (\ref{15} are
our main results, giving 
exact functional forms of a Wigner surmise type structure
for the universal spacing probabilities
$p_1(s)$, $p_2(s)$
and $p_4(s) - 2 p_1(2s)$. These expressions are well suited to the
numerical tabulations of the $p_\beta(s)$, 
or the generation of power series expansions thereof, although that is not our
concern here (accurate tabulations can be found in \cite{Ha92}).

Let us now turn to the derivation of (\ref{bw1}), (\ref{bw2})
and (\ref{15}. Consider
for example (\ref{bw2}) (the derivation of (\ref{bw1})
and (\ref{15}) is similar; also the derivation of (\ref{bw1}) will 
be presented as part of a forthcoming publication \cite{WF99b}).
The key ingredient is the fact from \cite{CS93} that the second
order second degree equation
\begin{equation}\label{c1}
x^2(y'')^2 = - 4(y')^2(xy' - y) + A_2(xy' - y) + A_3y' + A_4
\end{equation}
is solved in terms of a particular Painlev\'e V transcendent $u$
satisfying
\begin{equation}\label{c1'}
u'' = \Big ( {1 \over 2u} + {1 \over u - 1} \Big ) (u')^2 -
{u' \over x} + {(u-1)^2 \over x^2}\Big (\alpha u + {\beta \over u} \Big )
+ {\gamma u \over x}.
\end{equation}
This is the Painlev\'e V equation with $\delta = 0$ in standard notation.
As an aside we remark that it is known  \cite{CS93} that (\ref{c1'}) 
can always be solved in terms of a Painlev\'e III
transcedent. The relationship between (\ref{c1}) and (\ref{c1'}) is via
the formulas
\begin{eqnarray}
y & = & {1 \over 4u} \Big ( {xu' \over u - 1} - u \Big )^2 -
{1 \over 4}(1 - \sqrt{2 \alpha})^2(u-1) - {\beta \over 2} {u-1 \over u}
+ {\gamma x \over 4} {u+1 \over u - 1} \label{c1a} \\
y' & = & - {x \over 4u(u-1)} \Big (
u' - \sqrt{2\alpha} {u(u-1) \over x} \Big )^2 -
{\beta \over 2x} {u-1 \over u} - {\gamma \over 4} \label{c1b}  \\
A_2 = {\gamma^2 \over 4}, & &  
A_3 = \gamma \Big ( \beta + {1 \over 2} (1 - \sqrt{2 \alpha})^2 \Big ), 
\qquad
A_4 = {\gamma^2 \over 4}  \Big (  - \beta + {1 \over 2} (1 - 
\sqrt{2 \alpha})^2 \Big ). \label{c2}
\end{eqnarray}

Now it is easy to see that (\ref{c1}) reduces to (\ref{nw0}) if we write
\begin{equation}\label{h1}
x \mapsto s, \qquad y \mapsto - (\sigma_{B} - {1 \over 8} s - {1 \over 16}
\Big )
\end{equation}
\begin{equation}\label{h2}
A_2 = {1 \over 16}, \qquad A_3 = - {1 \over 16}, \qquad A_4 = {1 \over 128}.
\end{equation}
Substituting (\ref{h2}) in (\ref{c2}) gives
\begin{equation}\label{h3}
\sqrt{2 \alpha} = 1, \qquad \beta = - {1 \over 8}, \qquad
\gamma = {1 \over 2},
\end{equation}
while use of (\ref{h1}) and (\ref{h3}) in (\ref{c1a}) and (\ref{c1b})
allows us to deduce
\begin{equation}\label{h4}
{\sigma_B' \over \sigma_B} = - {u-1 \over s}.
\end{equation}
Furthermore we observe from (\ref{c1a}), (\ref{c1b}) with the substitutions
(\ref{h1}) and (\ref{h2}) that
\begin{equation}\label{h5}
\sigma_{B} + (u-1) + {1 \over 2} =: \tilde{\sigma}_{B}
\end{equation}
is also of the form (\ref{c1a}), (\ref{c1b}) but with
\begin{equation}\label{i1}
x \mapsto s, \quad
y \mapsto - ( \tilde{\sigma}_{B} - {1 \over 8}s - {9 \over 16} ), \quad
\sqrt{2 \alpha} = -1, \quad
 \beta = - {1 \over 8}, \quad \gamma = {1 \over 2}.
\end{equation}
These last three values substituted in (\ref{h2}) gives
$A_2 = {1 \over 16}$, $A_3 = {15 \over 16}$, $A_4 = {17 \over 128}$, and
these values together with the first two identifications in (\ref{i1})
substituted in (\ref{c1}) give the differential equation (\ref{nw3}).
The equation (\ref{a2}) follows from (\ref{h4}) and (\ref{h5}), together
with the facts deducible from (\ref{nw0'}) that $\tilde{\sigma}_{B}
\sim 0$ as $s \to 0$, while $\pi \sigma_{B}(s)/ s^{1/2} \sim 1$
and the boundary condition (\ref{bca}) follows similarly. 

We remark that the same procedure starting with $\tilde{\sigma}_{B}$
instead of $\sigma_{B}$ does not lead to a simple formula analogous
to (\ref{h4}), so we cannot expect the derivative of the RHS of (\ref{a2})
to simplify; the results (\ref{bw1}), (\ref{bw2}) and (\ref{15})
appear to be the
simplest functional forms possible. We emphasize that the structure of these
exact functional forms for $p_1(s)$, $p_2(s)$ and $p_4(s) - 2p_1(2s)$
are of the Wigner surmise type $a(s) e^{-b(s)}$, where instead of $a(s)$ and
$b(s)$ being simple power functions as in the approximation of Wigner,
$a(s)$ and $b(s)$ are Painlev\'e transcendents. 

\subsection*{Acknowledgements}
This work was supported by the Australian Research Council.


\end{document}